\documentclass[aip,jcp,reprint,amsmath,amssymb,floatfix,citeautoscript,superscriptaddress]{revtex4-1}

\usepackage{amsmath}
\usepackage{graphicx}
\usepackage{amssymb}
\usepackage{amsthm}
\usepackage{bm}
\usepackage[breaklinks=true,colorlinks=true,linkcolor=blue,citecolor=blue,urlcolor=blue]{hyperref}

\usepackage{multirow}
\usepackage{times}
\usepackage{txfonts}
\usepackage{xfrac}

\usepackage{footnote}
\usepackage{afterpage}

\usepackage{xcolor}

\begin{document}


\title{
Modeling Single Molecule Junction Mechanics as a Probe of Interface Bonding}

\author{Mark S. Hybertsen}
\email{mhyberts@bnl.gov}
\affiliation{Center for Functional Nanomaterials, Brookhaven National Laboratory, Upton, New York 11973-5000, USA}

\begin{abstract}
Using the atomic force microscope based break junction approach, 
applicable to metal point contacts and single molecule junctions,
measurements can be repeated thousands of times
resulting in rich data sets characterizing the properties
of an ensemble of nanoscale junction structures.
This paper focuses on the relationship between the measured
force extension characteristics including bond rupture
and the properties of the interface bonds in the junction.
A set of exemplary model junction structures have been
analyzed using density functional theory based calculations
to simulate the adiabatic potential surface that governs the junction elongation.
The junction structures include representative molecules
that bond to the electrodes through amine, methylsulfide and pyridine links.
The force extension characteristics are shown to be most effectively
analyzed in a scaled form with maximum sustainable force and
the distance between the force zero and force maximum as scale factors.
Widely used, two parameter models for chemical bond potential energy versus bond length
are found to be nearly identical in scaled form.
Furthermore, they fit well to the present calculations of N-Au and S-Au
donor-acceptor bonds, provided no other degrees of freedom are allowed to relax.
Examination of the reduced problem of a single interface, 
but including relaxation of atoms proximal to the interface bond,
shows that a single-bond potential form renormalized by an effective harmonic potential in series
fits well to the calculated results.
This allows relatively accurate extraction of the interface bond energy.
Analysis of full junction models show cooperative effects that go beyond the 
\textcolor{black}{
mechanical
}
series inclusion of the second bond in the junction, the spectator bond that does not rupture.
Calculations for a series of diaminoalkanes as a function of molecule length
indicate that the most important cooperative effect is due to the interactions
between the dipoles induced by the donor-acceptor bond formation at the junction interfaces.
The force extension characteristic of longer molecules such as diaminooctane,
where the dipole interaction effects drop to a negligible level,
accurately fit to the renormalized single-bond potential form.
The results suggest that measured force extension characteristics
for single molecule junctions could be analyzed with a modified potential form 
that accounts for the energy stored in deformable mechanical components in series.
\end{abstract}

\maketitle

\section{Introduction}

Mapping the potential surface that describes an individual chemical bond
represents a fundamental challenge,
all the more so when the bond in question
is formed at a metal-organic interface in a nanostructured system.
Nanostructured, heterogeneos materials are
ubiquitious in applications, such as catalysis.
In that context, understanding the diversity of binding sites
for molecules on the exposed surfaces of such materials 
and the relative stability of key bonds is of central importance.
However, quantitative measurement of individual, interface bonding characteristics
presents significant challenges.

Several approaches, based on the atomic force microscope (AFM) \cite{Giessibl03},
have emerged to provide different and complimentary measurements
probing the potential surface in bond formation, or the inverse process, bond rupture.
\textcolor{black}{
These studies can fall under the broader umbrella of mechanochemistry
in which mechanical forces are utilized to drive chemical processes.
\cite{Beyer05}
Recent research in this field
has been reanimated by the discovery that specific chemical moieties can be introduced into the
system to direct the energy supplied by mechanical means to perform controlled chemical bond
changes.
\cite{Hickenboth07, Davis09}
}

The AFM is a highly versatile scanned probe technique to measure surface topography.
Recent research based on controlled tip functionalization and 
more extensive mapping of the force versus distance from
the surface, but generally in non-contact mode, 
can reveal chemically
specific information \cite{Lantz01, Sugimoto07, Albers09, Welker12,DeOteyza13}.
In a different mode, the AFM is extensively used as a molecular force probe 
to study soft and biological molecules \cite{Muller08, Neuman08}.
Through contact with a sample surface, the probe tip adheres 
to a target species on the surface.
As the probe tip is withdrawn, a one-dimensional force versus elongation
characteristic is measured.
The mechanical response of soft and biological molecules
can be complex and details are outside the scope of this study.
However, early experiments along these lines were also able to isolate
features specifically attributable to rupture of covalent chemical bonds,
assigning a distinctive rupture force (of order nN) to different bonds \cite{Grandbois99}.
In a different realization of this concept, scanned and molecular force probe capabilities 
were combined;
the adhesion of a 3,4,9,10-perylene-teracarboxylic-dianhydride to the Au(111)
surface was measured through selective contact to a single molecule
and measuring the force-extension characteristic as it was lifted off the surface \cite{Wagner12}.

\begin{figure}[b]
\centering
\includegraphics[width=3.15in]{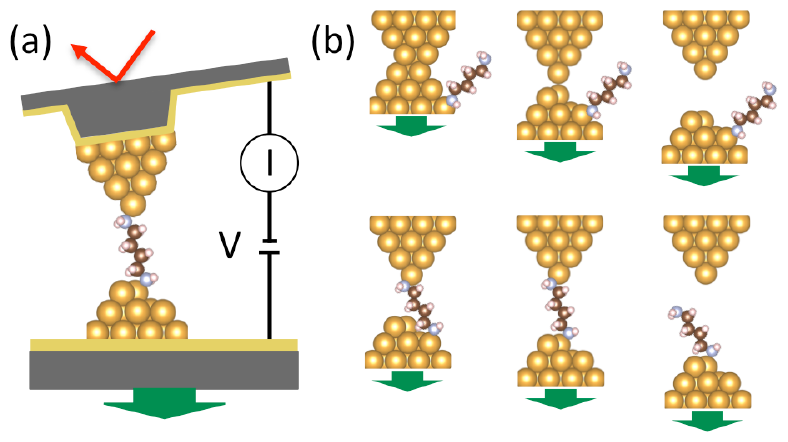}
\caption{
\textcolor{black}{
Break junction approach to simultaneously measure nanoscale junction conductance and force
as the junction is being pulled apart (AFM-BJ). (a) Schematic view of the set-up highlighting a junction
formed from diaminobutane. (b) Sketches illustrating key steps in the evolution of the junction during
elongation from a structure in which there is a nanoscale metal-metal contact through the stage when that
contact ruptures and a molecule bonds in the nanoscale gap to form a single molecule junction and finally
the rupture of one of the link bonds between the molecule and the electrodes. Representative traces of
measured conductance and force versus elongation appear in prior publications.
\cite{Frei11, Aradhya12, Aradhya14, Hybertsen16}
}
}
\label{fig:BJSchematic}
\end{figure}

Nanoscale junctions represent a different category of systems
in which the AFM is used as a molecular force probe.
Soon after the initial observation of conductance quantization
in metal atomic-scale junctions \cite{Olesen94},
mechanical characteristics of Au junctions
were measured simultaneously with electrical characteristics
as the junction was elongated under stress  \cite{Rubio96}.
\textcolor{black}{
Sketches illustrate
the method in Fig. \ref{fig:BJSchematic}.
}
A series of elastic and plastic deformations where observed
with correlation between the jumps in force and the changes in junction conductance.
The junctions consistently ruptured at a specific applied force, 
$\Delta F = 1.5 \pm 0.2~nN$.
Interestingly, the same characteristic rupture force was also observed in the regime,
achieved at low temperature and vacuum conditions,
in which a mono-atomic gold wire was formed during elongation,
extending to more than 1 nm prior to rupture \cite{RubioBollinger01}.
The details of the atomic structure in these measurements are generally not known.
The atomic-scale junction is formed \textit{in situ} as the Au contact area necks down under stress,
a process that generally follows a distinct path for each realization.
The quoted force in this experiment was the most probable 
from a histogram formed from order 200 individual measurements.

At the point when the metal point contact ruptures,
there is a sub-nm scale space that opens at a controlled rate
as the tip is systematically withdrawn.
In the presence of target molecules terminated by appropriate link groups,
either previously deposited on the electrodes or disolved in a solvent surrounding the tips,
small molecule bridges between these electrodes can form
single or few molecule circuit elements
\textcolor{black}{
(Fig. \ref{fig:BJSchematic}).
}
This approach to a mechanical break junction allows repeated
formation of target, single molecule junctions and characterization
of their specific conducting characteristics \cite{Xu03a}.
Harnessing selective chemistry, \textit{e.g.}, link groups such as amines that
selectively bond to undercoordinated Au atoms on the tips, 
results in highly reproducible conductance signatures \cite{Venkataraman06a}
and enables the direct correlation 
of conductance characteristics with molecular structure \cite{Venkataraman06b}.
This technique, among others, has led to significant progress
in understanding electronic transport through nanoscale molecular junctions,
including conductance phenomena with no classical analogue 
\cite{McCreery09, Aradhya13b, Nichols15, Hybertsen16}.

When this mechanical break junction approach is combined with an AFM set-up (AFM-BJ),
as it was in prior work on metal point contacts
\textcolor{black}{
and illustrated in Fig. \ref{fig:BJSchematic},
}
the electronic and mechanical characteristics of the single molecule junctions
can be simultaneously measured as function of elongation \cite{Xu03b, Frei11}.
In particular, the rupture force associated to the organo-metallic interface bond
breaking can be quantitatively correlated with changes in the bond characteristics,
as dictated by the molecular structure, \textit{e.g.}, 
distinguishing amine terminated alkanes, diaminobenzene
and 4,4'-bipyridine \cite{Frei11}.
In the measurement protocol where each iteration of the measurement
includes the fresh rupture of the electrode metal contact,
a new pair of metal tips naturally emerge with a nominally different atomic-scale
structure each time.
For each molecule measured, 
a database of thousands of individual junction measurements is stored,
\textit{i.e.}, conductance and force versus elongation.
The reduction of these large data sets to histograms of target, measured quantities,
such as linear response conductance and rupture force, 
support robust, quantitative analysis of trends
and comparisons to theory or simulation of representative examples.
However, as recently emphasized \cite{Hybertsen16},
the large database of individual junction data represents an opportunity.
In particular, behind the histograms, there is a wealth of data probing
the way changes in structure at the atomic scale in these nanoscale systems
affects such fundamental quantitities as the interface organo-metallic bond strength.

In the analysis of force extension characteristics, theory and simulation 
has contributed a great deal of 
\textcolor{black}{
insight.
\cite{RibasArino12}
}
Early simulations of metal-metal interactions during tip indentation, adhesion
and subsequent withdrawal and rupture of the contact, based on molecular dynamics,
highlighted a sequence of key plastic (inelastic) processes, 
particularly the necking down of the contact area to the atomic scale \cite{Landman90}.
The subsequent initial measurement of conductance quantization
in metal point contacts was accompanied by constructive simulations \cite{Olesen94}.
Atomic-scale processes occuring during elongation of metal
contacts under stress has now been widely studied 
with force-field based 
\textcolor{black}{
methods,
\cite{Sorensen98, RubioBollinger01, Bahn01, Dreher05, Pauly06, Pu10, French11, Wang11, Sabater12, French13} 
}
using Density Functional Theory (DFT) to calculate the energy \cite{Jelinek08, Thiess08, Tavazza09}
or a combination of both methods \cite{DaSilva01, DaSilva04}.
Two distinct approaches have typically been used: Driven molecular dynamics
or calculation of an adiabatic potential surface to describe the evolution of the system.
The molecular dynamics approach has the advantage of including inelastic events
in a natural way, albeit with significant limitations on the time scales
and for rates of junction evolution that are many orders of magnitude faster than experiments.
Also, constructive calculation of the force generally requires substantial extra effort,
although simpler, approximate approaches have been employed \cite{Finbow97}.
On the other hand, the force is trivially obtained from the adiabatic energy surface,
but the scope of structures and processes that can be explored is more limited.
Turning to the case of organo-metallic interfaces,
a key, early insight to the complexity of the thiolate-Au system
emerged from DFT-based molecular dynamics simulations of the evolution under stress
of a small organic molecule bonded to a model for the Au(111) surface via a thiolate linkage \cite{Kruger02}.
Instead of simply observing the stretching and rupture of the S-Au surface bond,
a much more complex sequence of inelastic events resulted in a chain of Au atoms
being pulled from the surface prior to a rupture event in which the Au atomic chain broke,
leaving three Au atoms bonded to the ethylthiolate.
For the single molecule junctions,
empirical force field \cite{Wang15} 
\textcolor{black}{
and DFT-based
\cite{Paulsson09, Strange10}
molecular dynamics simulations have been used,
as well as DFT-based calculations of the adiabatic potential landscape.
\cite{Stadler05, Romaner06, Batista07, Li07, Hoft08, Qi09}
}

\begin{figure}[b]
\centering
\includegraphics[width=3.0in]{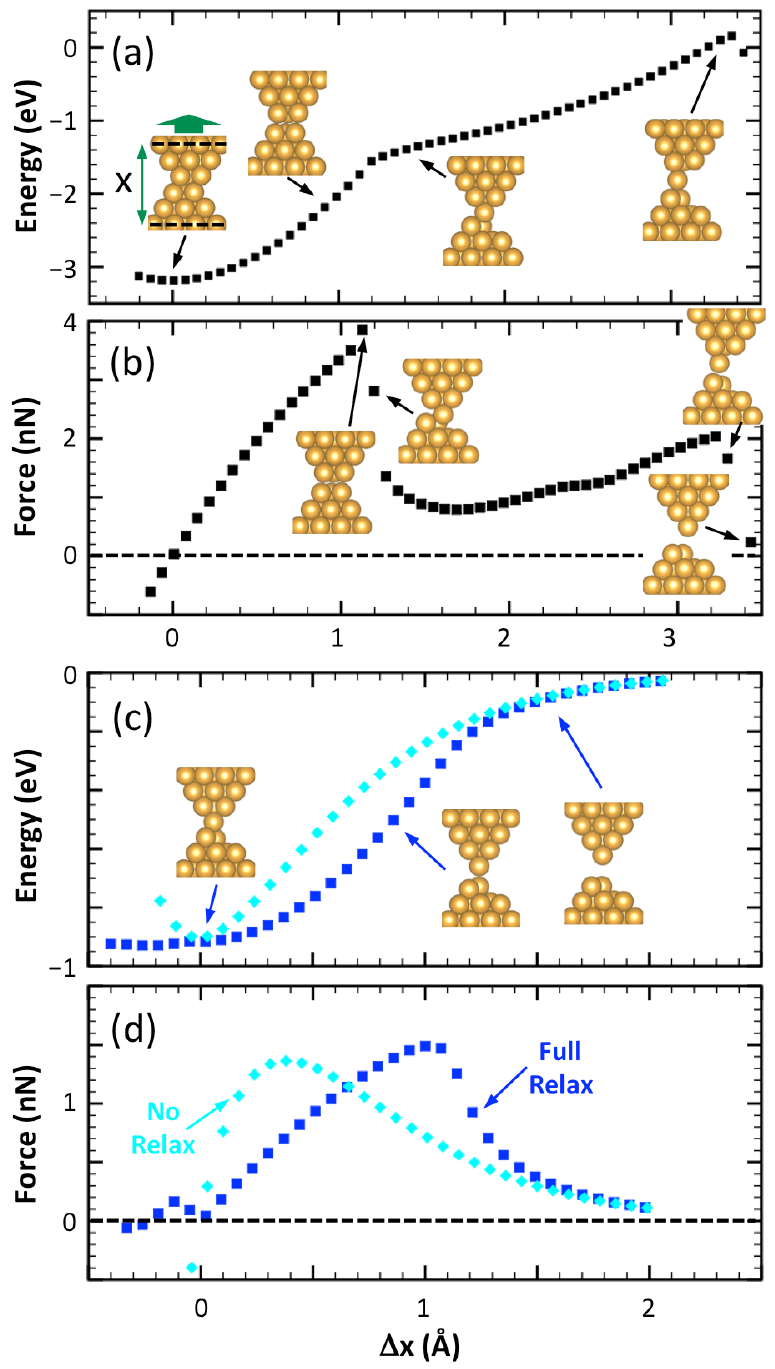}
\caption{
Simulation of Au point contact evolution under stress
using DFT-based calculations for energy and optimized structures
to determine an adiabatic potential surface.
(a, b) Energy and applied force 
\textcolor{black}{
($dU/dx$)
}
as a function of elongation 
for an example starting from a few atom contact area
showing initial, elastic evolution, the plastic cross-over to a potential surface
for a single atom contact area, its elastic evolution and final rupture of the contact.
(c, d) Energy and applied force for the evolution under stress
of a single atom contact formed from the same tip structures, 
but with slightly different alignment.
\textcolor{black}{
Insets show structural models of snapshots from several points on each potential surface.
The first inset in (a) illustrates the model set-up, with the top and bottom planes of Au atoms
held rigid at a fixed distance $x$.
Elongation corresponds to incrementing $x$ in small steps, allowing all other atoms to fully relax at each step.
For the results in (c) and (d) labeled 'No Relax', 
all the atoms in the two tip structures visualized in the final inset to (c) are held fixed for each step of elongation.
}
}
\label{fig:AuContact}
\end{figure}

In prior research, I developed and implemented approaches to simulate junction force extension
characteristics using a DFT-based approach
to compute an adiabatic potential surface \cite{Kamenetska09, Frei11, Frei12, Aradhya12, Aradhya14, Hybertsen16}.
As an example in Fig. \ref{fig:AuContact}, 
consider the basic features of an Au point contact evolution under stress,
here modeled with finite Au clusters to represent each tip of the electrodes.
Details of the methods used here are described in the body of this paper.
In general, as the junction is elongated, there are portions of
the junction extension that are nominally elastic 
(e.g., from 0 to 1 $\AA$ in Figs. \ref{fig:AuContact}a and b).
Then the atoms near the contact region rearrange in a plastic event (near 1.2 $\AA$).
and the system enters into a new, 
relatively stable structure that stretches elastically (from 1.4 to 3.4 $\AA$).
Both events involve a discontinuity in the structure where the system jumps
from one potential surface to another.
Before and after snapshots are visualized in Fig. \ref{fig:AuContact}b.
With a slightly different alignment of the gold tips to form
an atomic scale contact, that segment of the potential surface exhibits 
a single, smooth evolution under stress (Figs. \ref{fig:AuContact}c, d).
The force passes through a point of maximum sustainable force (about 1.5 nN)
and smoothly goes to zero as the gap between the two tips opens up.

The observed evolution of the junciton is not 
solely determined by the 
\textcolor{black}{
elongation
}
of the local bonds that bridge the contact point.
To isolate the role of stretching the Au-Au distance specifically at the contact from
the deformation of the structure nearby, 
\textcolor{black}{
further constraints were applied. 
Referring to the insets in Fig. \ref{fig:AuContact}c, the entire structure of each model tip
was held rigid. 
As the relative distance between the tip structures was varied, 
no atomic relaxation was allowed. 
Thus, the only bond distance that was allowed to change was the Au-Au separation at
the contact point. 
Relative to the unconstrained case, the slope (stiffness) is larger, 
the maximum sustainable force is slightly lower 
and the distance traversed to that maximum is smaller. 
Beyond the point of elongation where the force is maximum, the force drops more slowly as a function of elongation. 
Thus, in the fully relaxed calculation, representative of the realistic case where the tips are deformable,
stress-induced changes of the near-by tip structure during initial elongation store energy that is then released
as the force drops beyond the maximum.
}

\textcolor{black}{
A critical issue when interpreting measurements of bond rupture force is how it correlates with
the maximum sustainable force of a specific bond. 
As discussed in detail in a recent review,
\cite{RibasArino12}
the full mechanical system of the measurement must be considered, 
including the boundary conditions
determined by the way the measurements are executed. 
For example, in the case of force spectroscopy
in soft materials, the system is subject to a constant force boundary condition. 
That force is then ramped up at a specific loading rate. 
The constant applied force corresponds to a linear
potential that must be combined with the potential 
representing a specific bond to model the total mechanical system. 
In this case, the system is inherently unstable along the direction of applied force.
The system always has a lower total energy with the bond ruptured. 
A barrier separates the system configuration with the bond intact from that with the bond ruptured,
but that barrier reduces to zero as the applied force approaches the maximum sustainable
force for the bond. 
Closely related to the Kramers model for rate of escape from a bound state in an applied field,
\cite{Kramers40, Hanggi90}
the resulting distribution of measured rupture forces reflects stochastic processes that carry the
system over the barrier, including dependence on the rate of loading.
\cite{Evans01}
In the case of the AFM-BJ measurements that are the focus of this study, 
the system is subject to elongation at a constant rate; 
separation is the control variable and force is measured. 
At least for analysis purposes, the balance of the system, 
including the AFM cantilever, can be represented by a harmonic potential.
In this case, the inherent stability of the system along the direction of pulling is contingent on a
stiffness criterion.
\cite{Friddle08b}
As discussed elsewhere,
\cite{Aradhya14, Hybertsen16}
the AFM-BJ experiments can be carried out with a stiff cantilever 
such that bistability in the one-dimensional potential surface along the stretching
direction is avoided. 
In that case, the full force extension characteristic, illustrated in the example
in Figs. \ref{fig:AuContact}c and d, could be accessible, 
including the region beyond the force maximum. 
Of course, as the bond is elongated, the energy of the system is raised 
and barriers for fluctuations of other degrees of freedom 
such as rotation of the molecular backbone away from the direction of
elongation become smaller. 
Detailed models of the bond
rupture processes that occur in the AFM-BJ measurements remain to be understood.
}

In this paper, I focus on understanding the essential factors that describe
the interplay between the characteristics of the individual
organo-metallic link bonds at
the junction interfaces and the overall junction evolution under stress.
I will not consider further the stochastic 
\textcolor{black}{
processes that govern bond rupture in the AFM-BJ set-up,
although I will briefly analyze the conditions for inherent stability along pulling direction.
}
For the purpose of elucidating the force extension characteristics, 
I use the data generated from a series of DFT-based simulations,
such as those illustrated in Fig. \ref{fig:AuContact} for the special case of an Au point contact.
The long-term goal is to support quantitative analysis
of individual, measured force extension data to extract link bond energy.
The results presented here extend prior, initial work towards this goal 
in which a simple model form was developed to fit data
and scaling behavior of junction force extension characteristics was introduced \cite{Aradhya14, Hybertsen16}.
My approach here starts with an assessment of widely used, simple models to
describe individual bonds.
I then systematically include the relaxation
of the near-by atoms and finally cooperative effects that affect the full junction evolution.
\textcolor{black}{
It will emerge that detailed information near and beyond the
force maximum are important for disentangling the individual bond characteristics from those
of the junction as a whole. 
Also, determining the bond energy requires capturing the essential
characteristics from the force maximum through the asymptotic region, 
at least as represented by
the form of the model and the deduced parameterization of it.
}

In the next section, analytical models used in this work are discussed.
In Sect. III, the DFT-based methods are described.
The main results and discussion appear in Sect. IV, 
followed by concluding remarks in Sect. V.

\section{Analytical Models}

\begin{figure}[b]
\centering
\includegraphics[width=3.1in]{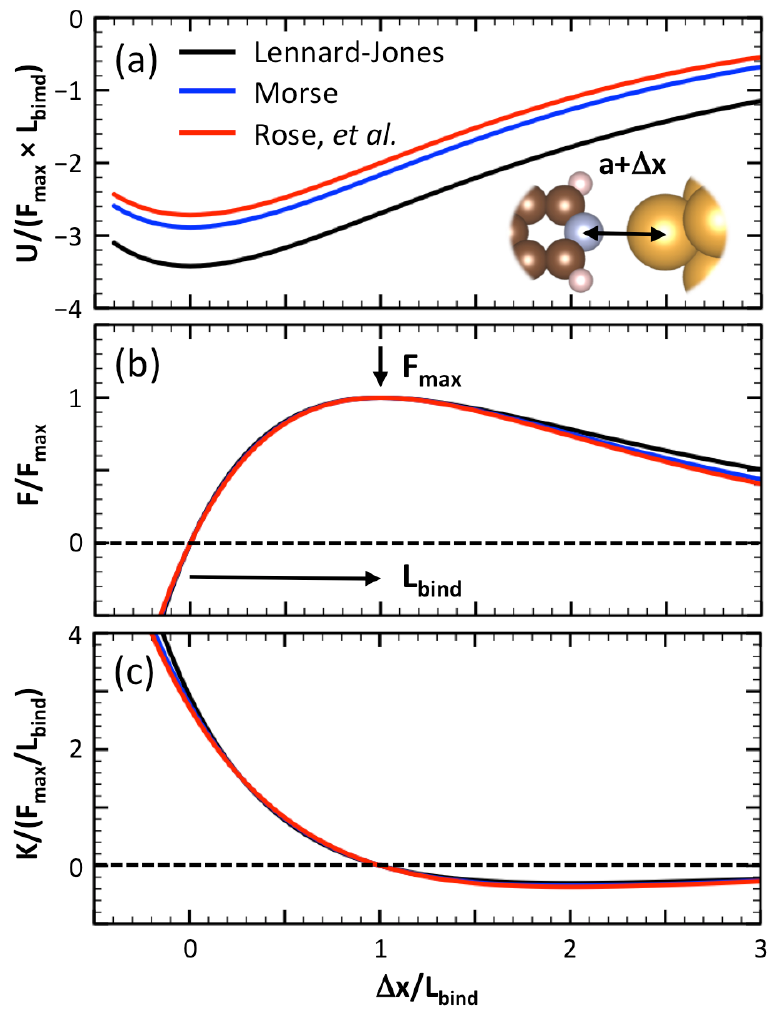}
\caption{
(a) Three commonly used models to represent the potential energy surface
of a single chemical bond, such as the exemplary N-Au bond from a
model for a molecule-electrode interface illustrated in the inset. 
\textcolor{black}{
Bond length change $\Delta x$ is measured relative to the equilibrium bond length $a$.
}
(b) Force versus change in bond length.  
(c) Corresponding stiffness versus change in bond length.
In all panels, the distance,
energy, force and stiffness are scaled, as indicated, using the parameters
of maximum sustained force ($F_{max}$) and distance
from the potential minimum to the inflection ($L_{bind}$) as defined in (b).
}
\label{fig:BondModels}
\end{figure}

A minimal model for the bond length dependence 
of the potential surface describing a chemical bond 
requires an energy scale $U_0$ and a distance scale $d$.
One popular choice for solids is the so-called `universal'
model proposed by Rose, Ferrante and Smith \cite{Rose81},
here displayed as a function of the change in bond length relative
to that at the minimum energy ($\Delta x$):
\begin{equation}
U(\Delta x) = -U_0 \left( 1+ \Delta x /d \right) e^{-\Delta x / d}. 
\label{eq:RoseU}
\end{equation}
The mechanical force required to stretch or compress the bond
is given by the first derivative and the corresponding bond stiffness
by the second derivative:
\begin{eqnarray}
F(\Delta x) = U'(\Delta x) & = & (U_0 /d) (\Delta x /d) e^{-\Delta x / d},
\label{eq:RoseF}\\
K(\Delta x) = U''(\Delta x) & = & (U_0 /d^2) \left( 1- \Delta x /d \right) e^{-\Delta x / d}.
\label{eq:RoseK}
\end{eqnarray}
Here, a positive force will correspond to the physical system
under tension and I will resolve any ambiguity regarding
how the force is applied in context.
For the example of a single N-Au bond illustrated
in the inset to Fig. \ref{fig:BondModels}a,
that isolated bond under tension ($\Delta x > 0$) corresponds
to a force applied to the Au atom to the right
and an equal and opposite force to the N atom to the left.
Equation (\ref{eq:RoseF}) implies that $F=0$ for $\Delta x=0$.
The force and potential also go to zero at large separation.
The force is nonlinear outside the region near $\Delta x=0$ and
there is a maximum sustainable force for this bond
defined by the inflection in the potential or the point at which
the stiffness $K(\Delta x)$ in Eq. (\ref{eq:RoseK}) is zero.
This occurs for $\Delta x = L_{bind} = d$ and for which $F_{max} = U_0 / ed$.

\begin{center}
\begin{table*}[t]
\begin{tabular*}{0.9\linewidth}{@{\extracolsep{\fill}} l c c c c c}
\hline
\hline
  & $F_{max}$ & $L_{bind}$ & $L_{bind}K(0)/F_{max}$ & $U(0)/F_{max}L_{bind}$ & $U(L_{bind})/U(0)$ \\
\hline
Lennard-Jones & $12\beta U_0/d$ & $(\alpha -1)d$ & $2.9090$ & $-3.4207$ & $0.7870$\\
Morse & $U_0/2d$ & $ln(2)d$ & $2.7726$ & $-2.8854$ & $0.75$ \\
Rose, \textit{et al.} & $U_0 /ed$ & $d$ & $2.7183$ & $-2.7183$ & $0.7358$ \\
\hline
\hline
\end{tabular*}
\caption{
Summary of the characteristics of three commonly used models for chemical bond potential landscape.
The parameters appearing in the Lennard-Jones expressions are $\alpha = (13/7)^{(1/6)} =1.1087$
and $\beta = (7/13)^{(7/6)}-(7/13)^{(13/6)} = 0.2242$.
}
\label{tab:BondModels}
\end{table*}
\end{center}

The primary focus here will be to understand force-extension characteristics.
The parameters $F_{max}$ and $L_{bind}$ are natural choices
to characterize the system.
Furthermore, in prior work, it was found that these parameters can
be used to rescale and collapse a substantial amount of experimental data \cite{Aradhya14, Hybertsen16}.
The corresponding measure for energy is of course $F_{max} \times L_{bind}$
and in those units $U(\Delta x =0) = -e$.
The scaled potential, force extension, and stiffness for the universal potential of Rose, \textit{et al.},
are plotted in Fig. \ref{fig:BondModels}.
In the same plot, two other widely used models are also shown:
the Morse potential \cite{Morse29}
\begin{equation}
U(\Delta x) = U_0 \left( e^{-2\Delta x /d} - 2 e^{-\Delta x / d} \right). 
\label{eq:MorseU}
\end{equation}
and the Lennard-Jones potential \cite{Jones24}
\begin{equation}
U(\Delta x) = U_0 \left( \frac {1} {(1+ \Delta x /d)^{12}} -  \frac {2} {(1+ \Delta x /d)^6} \right). 
\label{eq:LJU}
\end{equation}
Both expressions have been formulated in terms of
nominally the same variables used for the universal potential.
Straightforward analysis yields 
the expressions for $F_{max}$, $L_{bind}$ as well as scaled values for the stiffness and potential.
The results are collected in Table \ref{tab:BondModels}.

\textcolor{black}{
Of course, the simplest model potential with a force maximum 
would be a cubic polynomial. 
While it is a useful fit near the force maximum, 
it does not serve the present purpose as a global form 
since it diverges for large distance. 
Also it has a linear stiffness, 
quite different in detail from the models shown in Fig. \ref{fig:BondModels}c.
}

From the expressions for the $L_{bind}$ parameter characterizing the distance from
the potential minimum to the maximum sustainable force point (inflection in the potential),
it is clear that the distance parameter $d$ has a different physical meaning among
the different models.
\textcolor{black}{
In particular, in the Lennard-Jones model, the parameter $d$ is the equilibrium bond length,
while in the models of Rose, \textit{et al.}, and Morse, the equilibrium bond length does not explicitly enter.
}
However, once the models are constrained to give the same $F_{max}$ and $L_{bind}$, 
arguably conditions that should then indicate a very similar chemical bond,
then the other physical characteristics align more closely.
In particular, the bond stiffness near the equilibrium bond length only varies in a range of about $7\%$.
Correspondingly, the rescaled force-extension curves for the three models are nearly indistinguishable
from modest compression right through the maximum sustainable force (Fig. \ref{fig:BondModels}b).
They only become distinguishable beyond about $2 \times L_{bind}$.
There, the Lennard-Jones model in particular starts to deviate.
Correspondingly, the potential curves in Fig. \ref{fig:BondModels}a show
the slower decay with distance that must follow from the power law in that case,
as compared to the exponential dependence for the other two models.
Of course, the power law reflects dispersion interactions between separated atoms.
While this has a relatively small influence over the range of $L_{bind}$,
the integrated impact does lead to a deeper binding energy for the
same $F_{max}$ and $L_{bind}$.
All the models are designed to reference zero at large distance,
so this is reflected in the normalized potential at zero extension shown in Table \ref{tab:BondModels}.
One may assess the difference between the Lennard-Jones model
and the other two in this regard as on estimate for the impact of dispersion
interactions in this context ($20 - 25 \%$).

\begin{figure}[b]
\centering
\includegraphics[width=3.1in]{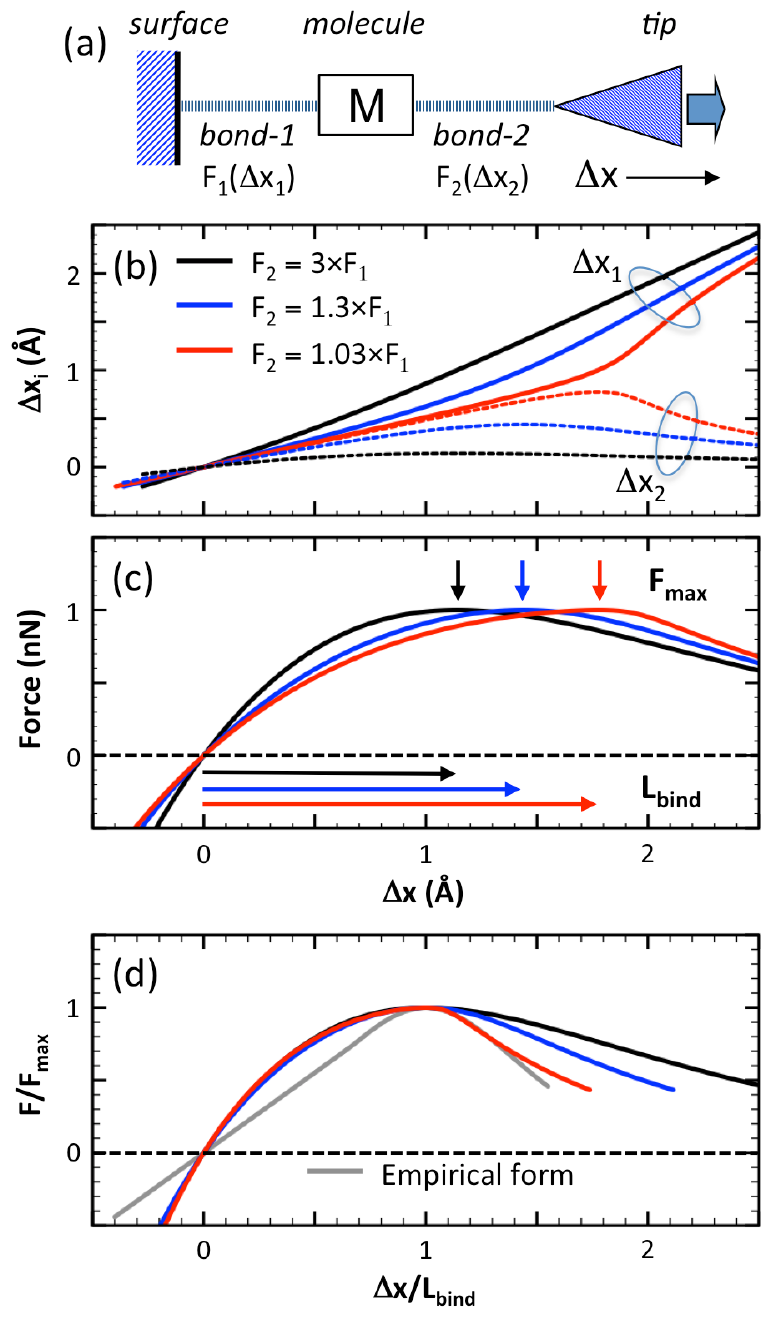}
\caption{
Mechanical model for a single molecule junction.
(a) Schematic representation of the essential mechanical components
of a single molecule junction formed in an atomic force microscope set-up. 
(b) Elongation of individual link bonds and (c) force versus net elongation
of the junction for three values of $F_{2,max}$. 
$F_{1,max} = 1~nN$ and $L_{1,bind} = L_{2,bind} = 1~\AA$ are used.
In (c), the net $L_{bind}$ for the junction and the corresponding position of $F_{max}$
are illustrated.
(d) The scaled force-extension curves for all three cases along with
an empirical functional form used in prior research to fit experimental data \cite{Aradhya14, Hybertsen16},
as discussed in the text.
}
\label{fig:ModelJunctions}
\end{figure}

Now consider a simplified model for a single molecule junction formed
and elongated in an AFM-BJ set-up, illustrated in Fig. \ref{fig:ModelJunctions}a.
For the experimental conditions envisioned here, 
the AFM cantilever is substantially stiffer than the microscopic junction.
The additional extension due to cantilever deflection can be safely neglected.
Turning to the microscopic junction, 
consider the case where the metal-organic links
form a donor-acceptor bond (e.g., amine N-Au).
These will be the softest structural components in the junction.
Then the model consists of two independent link bonds, each one
characterized by a force extension characteristic.
In static equilibrium, elongation of each of these bonds must
result in equal and oppositely directed forces on the central, static
molecular backbone.  In turn, the magnitude of this force
must equal that registered by the external cantilever deflection.

Each link bond is characterized by a maximum sustainable force.
Overall, assuming an adiabatic extension of the junction, the smallest
maximum sustainable force must control the force-extension
characteristic of the whole junction.
To be definite, assume $F_{1,max} < F_{2,max}$.
Then bond 1 controls the junction evolution and the junction
force extension curve can be determined using $\Delta x_1$ as the independent variable:
\begin{eqnarray}
\Delta x & = & \Delta x_1 + \Delta x_2 (\Delta x_1),
\label{eq:JunctionX}\\
U(\Delta x) & = & U_1 (\Delta x_1) + U_2 (\Delta x_2 (\Delta x_1)).
\label{eq:JunctionU}\\
F(\Delta x) & = & F_1 (\Delta x_1) = F_2 (\Delta x_2 (\Delta x_1)).
\label{eq:JunctionF}
\end{eqnarray}
Note that Eq. (\ref{eq:JunctionF}) denotes two bonds under the same degree
of tension or compression and the second equality determines
the response of bond 2: $\Delta x_2 (\Delta x_1)$.
Then the potential energy for the junction is given by Eq. (\ref{eq:JunctionU})
and it follows that $dU/d\Delta x$ gives $F(\Delta x)$ in Eq. (\ref{eq:JunctionF}).
The stiffness of the junction follows:
\begin{equation}
K(\Delta x) = \frac {dF} {d\Delta x} = \frac {K_1 (\Delta x_1)} {1+K_1 (\Delta x_1)/K_2 (\Delta x_2)}. 
\label{eq:JunctionK}
\end{equation}
Since bond 2 will never extend to its maximum sustainable force, $K_2 (\Delta x_2)$
will always be greater than zero.
Once bond 1 extends beyond $L_{1,bind}$, the stiffness $K_1 (\Delta x_1)$
will be negative.
In this one-dimensional model, the junction stability requires
$K_2 (\Delta x_2) > \left| K_1 (\Delta x_1) \right|$ in this region.
Bistability occurs when this condition is violated.
\textcolor{black}{
For illustration, consider the case where bond-1 is represented by
one of the model forms illustrated in Fig. \ref{fig:BondModels}.
Then the stability criterion compares $K_2 (\Delta x_2)$
to the most negative stiffness seen in  Fig. \ref{fig:BondModels}c near $\Delta x / L_{bind} =2$.
From this picture, it is natural to define a critical stiffness 
for any potential model: $K_{c} = \left| min( K_1 (\Delta x)) \right|$.
For the models considered here, 
quantitatively $K_c/(F_{max}/L_{bind})$ = -0.368 (-0.346, -0.302) for
the Rose, \textit{et al.}, (Morse, Lennard-Jones) potential.
The critical stiffness is much smaller than the stiffness near equilibrium (Table \ref{tab:BondModels}).
}

Figure \ref{fig:ModelJunctions}b displays numerical results for the bond extension in
three cases where the bonds are represented by the universal potential
of Rose, \textit{et al.}, with equal $L_{bind}$ but different ratios of $F_{max}$.
As the two link bonds become more similar, the second, stiffer bond
extends further, storing more energy prior to the maximum sustainable force.
As shown in Fig. \ref{fig:ModelJunctions}c, the junction force extension curve
exhibits lower stiffness and larger effective $L_{bind}$.
For extension beyond $L_{bind}$, the stiffer bond relaxes, returning energy
and the relative rate of extension is slower.
As seen, this results in a force-extension characteristic that is asymmetric
around the maximum sustainable force.
Scaled results, shown in Fig. \ref{fig:ModelJunctions}d, emphasize this point.
\textcolor{black}{
In all of these examples,
$K_2 (\Delta x_2) \gg K_{1,c}$, the stability criterion just introduced.
However, as $F_{1,max}$ approaches $F_{2,max}$, 
the drop in the model junction force extension after $F_{max}$ becomes steeper 
and there is a different instability in the
limit of a symmetric junction where both $K_1 (\Delta x_1)$ and $K_2 (\Delta x_2)$
are approaching zero together. 
Such symmetric junctions are unlikely in experiments, but examples have been encountered
in simulations described in Section IV that behave this way.
}

In prior work, analyzing both experimental data and DFT-based simulations,
it was found that junction force-extension characteristics remained harmonic
over a substantially larger fraction of the distance up to $L_{bind}$ \cite{Aradhya14, Hybertsen16}.
A simplified, two-parameter form was developed to fit that data.
The model potential consisted of a parabolic segment $U_0 + K \Delta x^2 /2$ and 
a logistic segment $-D/(1 + e^{(\Delta x - L_{bind})/r})$,
with appropriate boundary conditions described previously \cite{Aradhya14}.
This form fit a large body of experimental data from individual junction
force extension traces, although that data had a limited representation
in the region of extension beyond $L_{bind}$.
The fits to DFT-based simulations 
showed more deviation in that region beyond $L_{bind}$ \cite{Hybertsen16}.
For reference, the empirical form is plotted in Fig. \ref{fig:ModelJunctions}d.

\begin{figure}[b]
\centering
\includegraphics[width=3.1in]{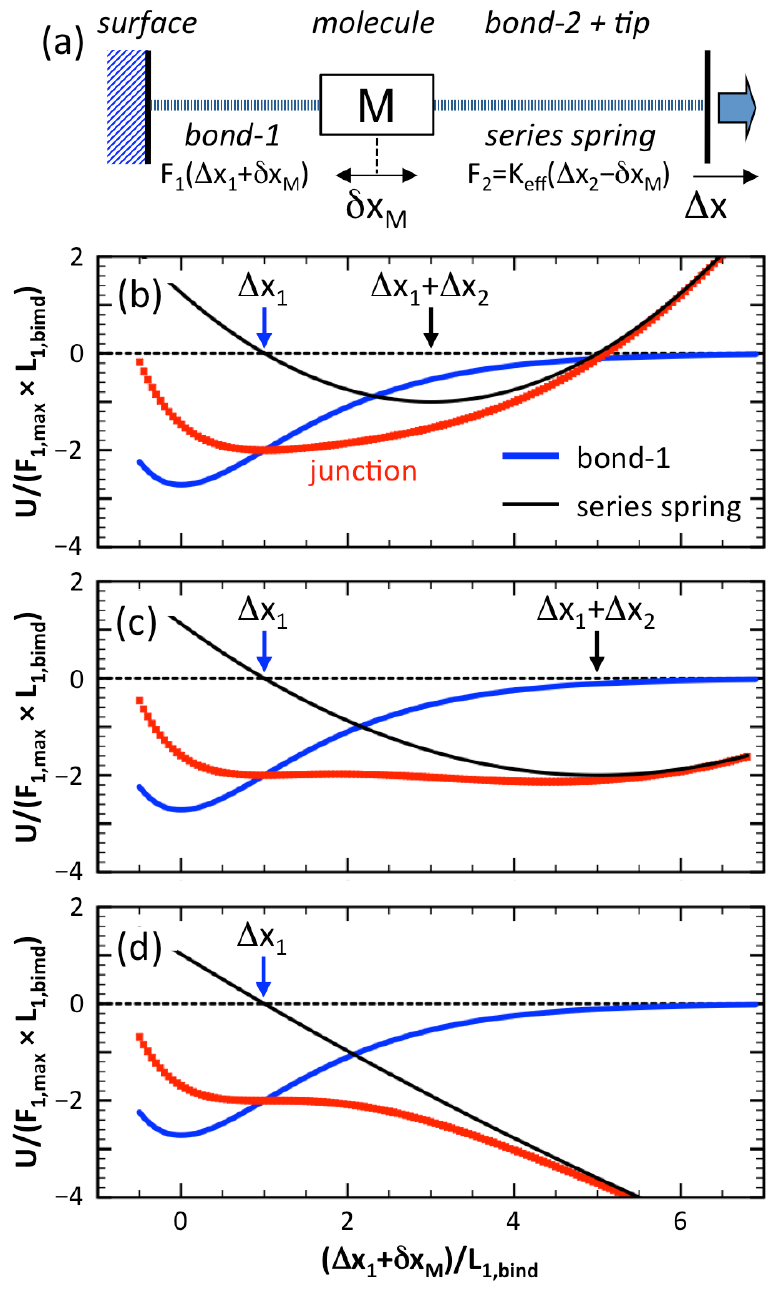}
\caption{
\textcolor{black}{
Simplified mechanical model for a single molecule junction for stability analysis. 
(a) With reference to Fig. \ref{fig:ModelJunctions}a, all deformable components of the junction outside bond-1, 
assumed to control rupture,
are replaced by a single, effective spring with stiffness $K_{eff}$. 
The internal displacement of the molecule
away from equilibrium is $\delta x_M$. 
The plots show the energy for bond-1, the series spring model and the total
system in renormalized units as a function of molecule internal displacement $\delta x_M$. 
The elongation is chosen so that $\Delta x_1 = L_{1,bind}$, the condition for maximum sustainable force of bond-1. Results for three different values of $K_{eff}$ (units of $F_{1,max}/L_{1,bind}$) 
are shown according the stability criterion: 
(b) $K_{eff} = 0.5 > K_{1,c}$; (c) $K_{eff} = 0.25 < K_{1,c}$;
and (d) $K_{eff} = 0.05 \ll K_{1,c}$.
}
}
\label{fig:JunctionStability}
\end{figure}

\textcolor{black}{
Physically, the more harmonic response observed in experiment suggests consideration of the
mechanical components of the junction in series with the bond that ruptures. 
In what is likely the more typical case where the
two metal-organic link bonds have different bond strength, 
assume that $F_{2,max}$ is sufficiently larger
than $F_{1,max}$ so that bond-2 does not appreciably depart from a harmonic response. 
Then Eq. \ref{eq:JunctionF} can be solved assuming
a constant $K_2$ and Eq. \ref{eq:JunctionX} yields:
\begin{equation}
\Delta x =  \Delta x_1 + F_1(\Delta x_1) /K_2.
\label{eq:HarmonicX}
\end{equation}
The deformation of each of the metal electrode tip structures in response to the applied stress
will also be sources of elongation in series. 
In addition there is the displacement of the AFM cantilever.
Assuming the response of each is elastic, then the mechanical analysis of the junction can be
envisioned as a series of springs. 
Maintaining the focus on the bond-1 that ruptures, the net result is a simple generalization
where $K_2$ is replaced by
\begin{equation}
\frac {1} {K_{eff}} = \frac {1} {K_2} + \frac {1} {K_3} +  \frac {1} {K_4} +  ...
\label{eq:SeriesK}
\end{equation}
leading to a simplified mechanical model for the junction, illustrated in Fig. \ref{fig:JunctionStability}a.
}
Characterizing the composite junction, $F_{max} = F_{1,max}$ flows
through, assuming static equilibrium,
but $L_{bind}$ is expanded, naturally:
\begin{equation}
L_{bind}/L_{1,bind} =  1 + (F_{1,max}/L_{1,bind}) /K_{eff}.
\label{eq:HarmonicL}
\end{equation}
For applications below, it is simplest to express $K_{eff}$
in dimensionless units that reference the parameters for the bond, $F_{1,max}/L_{1,bind}$.

\textcolor{black}{
Finally, returning to the question of inherent stability, 
the junction model illustrated in Fig. \ref{fig:JunctionStability}a 
captures all the deformable components of the junction, outside bond-1 that will rupture, 
as a composite, harmonic potential:
\begin{equation}
U_2(\Delta x_2) = U_{2,0} +\frac {1} {2} K_{eff} \Delta x_2^2.
\label{eq:EffHarmonU}
\end{equation}
These include the other bond, electrode deformation and the AFM cantilever. This model to
analyze stability, adapted here, has been used previously.
\cite{Friddle08b}
The measured elongation is the total $\Delta x$ as before 
and the internal position of the molecule backbone (assumed rigid) is determined
from static equilibrium, expressed through $\Delta x_1$ and $\Delta x_2$. 
Stability is then framed in terms of
the potential energy surface that governs the internal deviations of the molecule away from its
equilibrium position, $\delta x_M$:
\begin{equation}
U(\delta x_M) = U_1(\Delta x_1 + \delta x_M) + U_2(\Delta x_2(\Delta x_1) - \delta x_M ).
\label{eq:EffHarmK}
\end{equation}
Instability is signaled by the appearance of two minima for the position of the molecule in this potential. 
A maximum (or an inflection) must appear between these, 
indicated by a zero in the derived
force and a negative (zero) value of the stiffness at that position.
Focusing on the stiffness,
\begin{equation}
K(\delta x_M) = \frac {d^2U} {d \delta x_m^2} = K_1(\Delta x_1 + \delta x_M) + K_{eff} .
\label{eq:EffHarmonK}
\end{equation}
The stability criterion $K(\delta x_M) > 0$ is the same as that discussed
below Eq. \ref{eq:JunctionK}.
For an inherent instability to appear along the
direction of elongation in this model, the value of $K_{eff}$ must be smaller than $K_{1,c}$.
}

\textcolor{black}{
Stability in this model is further illustrated for several cases in Fig. \ref{fig:JunctionStability}. 
Stability is probed for the conditions of elongation
that place bond-1 at its maximum sustainable force. 
First $K_{eff}$ = $0.5F_{1,max}/L_{1,bind}$
is considered, a stiffness that exceeds the stability criterion. 
Then the corresponding equilibrium value of $\Delta x_2$ = $2L_{1,bind}$. 
To make visualizing the relationship between the system potential and the bond-1 potential easy, 
$U_{0,2}$ is chosen so that $U_2 (\delta x_M)= 0$ when $\delta x_M = 0$. 
Then for this value of $K_{eff}$, $U_{0, 2}$ = $-1 (F_1{,max} \times L_{1,bind})$.
As shown in Fig. \ref{fig:JunctionStability}b, 
the molecule is inherently stable at $\delta x_M = 0$, despite being poised at the
maximum sustainable force for bond-1. 
There is no inherent instability in this case. 
However, choosing $K_{eff}$ = $0.25F_{1,max}/L_{1,bind}$, 
below the stability requirement, gives a different result. 
In this case $\Delta x_2$ = $4L_{1,bind}$ and $U_{0,2}$ = $-2 (F_{1,max} \times L_{1,bind})$ is chosen. 
Now bistability appears, as illustrated in Fig. \ref{fig:JunctionStability}c. 
In fact, the system enery is minimized if the molecule jumps to $\Delta x_2$, 
with only a small barrier to overcome.
Under these conditions, the system becomes inherently unstable as the elongation approaches
the conditions of maximum sustainable force on bond-1. 
Bond rupture is determined by kinetic
process that carry the system over the barrier. 
Finally, the case with a very soft effective spring constant, 
$K_{eff}$ = $0.05F_{1,max}/L_{1,bind}$, is shown in Fig. \ref{fig:JunctionStability}d. 
Now $\Delta x_2$ = $20L_{1,bind}$ far exceeds the scale
of the single bond. 
The potential due to the applied force is now approaching linear (the limit
of a constant force typical of force spectroscopy experiments). Not only is the system inherently
bistable, but the barrier is approaching zero.
}

\textcolor{black}{
As discussed elsewhere,
\cite{Aradhya14, Hybertsen16}
through the choice of a stiff cantilever and the characteristics of
the balance of the junction, 
it is entirely feasible for $K_{eff}$ to be maintained at a value larger than
the inherent instability criterion $K_{1,c}$ in AFM-BJ experiments. 
Thus, in principle, it is possible to
probe the intrinsic force extension characteristic of interface bonds through the point of maximum
sustainable force and beyond. 
In practice, stochastic events, involving other degrees of freedom,
can and will intervene to limit the scope of the segment of the force extension characteristic
probed, but these are not tied to the force maximum as such.
}

\section{Numerical Methods}

The results presented here build on and extend prior work \cite{Kamenetska09, Frei11, Frei12, Aradhya12, Aradhya14, Hybertsen16}.
For completeness, the technical details are described here.

\begin{figure}[b]
\centering
\includegraphics[width=2.8in]{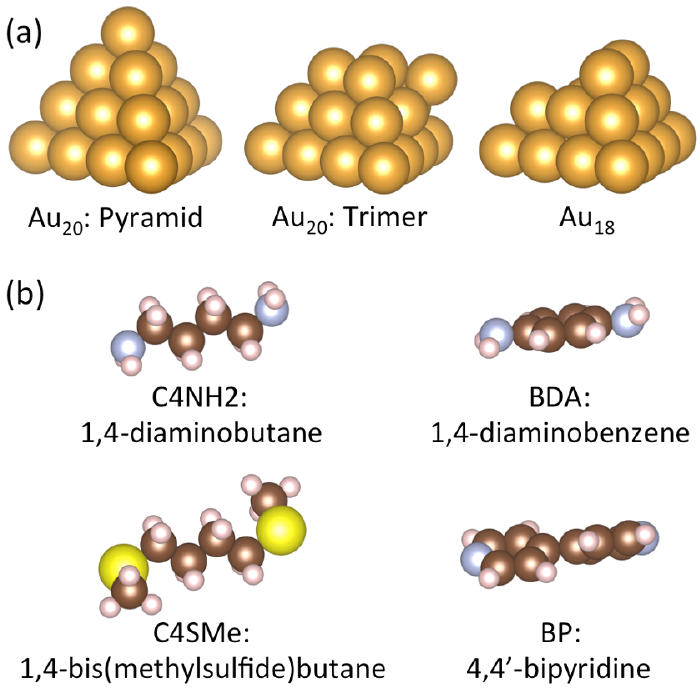}
\caption{
(a) Three models used to represent metal electrodes. 
(b) Models illustrating four molecules considered here.
}
\label{fig:ModelStructures}
\end{figure}

Since the junctions formed in the AFM-BJ method are known to bridge local asperities on the electrodes,
these are modeled by tip-like structures.
Specifically, each electrode is modeled by a pyramidal structure consisting of 20 atoms,
with 10 atoms in the base layer and exposing three (111) facets on the sides.
Periodic slab models for electrodes, used in some prior work, will not be further discussed here.
As illustrated in Fig. \ref{fig:ModelStructures}a,
one electrode is modeled by an ideal pyramid and the other by one in which the
top-most atom is displaced to a pyramid side.
A third model with 18 atoms is also shown. It was used as the second electrode
for the metal point contact simulations illustrated in Fig. \ref{fig:AuContact}.
The base layer of these tip structures is held fixed with atom separations taken
from the optimized, bulk metal DFT calculations.
Most of the junctions considered here involved Au contacts,
but some examples with Ag electrodes were also considered.
The molecules considered included amine, methylsulfide and pyridine link groups.
Exemplary molecular structures are visualized in Fig. \ref{fig:ModelStructures}b.
The use of two distinct electrode structures for the two junction contacts
is by design.
In the experiments, a highly symmetric junction should be an exceptional case.
Furthermore, as illustrated by the model calculations in Fig. \ref{fig:ModelJunctions},
junctions in which the individual bond force extension characteristics are identical
can lead to a singularity in the junction behavior near the force maximum.

The DFT-based total energy and structure optimization calculations were
performed with the Vienna \textit{ab initio} simulation package (VASP) \cite{Kresse96}.
\textcolor{black}{
VESTA was used to visualize structures.
\cite{Momma11}
In VASP, the
}
projector augmented wave approach was used \cite{Blochl94, Kresse99}.
This naturally includes scalar relativistic effects for heavy elements such as Au.
For the exchange-correlation functional, the generalized gradient approximation (GGA)
developed by Perdew, Burke and Ernzerhof (PBE) was employed \cite{Perdew96}.
Although dispersion interactions were approximately included in one prior study \cite{Aradhya12},
the force extension characteristics of specific chemical bonds is the primary focus here.
These are quantitatively affected by dispersion interactions, but to a minor extent.
The Kohn-Sham equations were solved with a basis set determined by a 400 eV cutoff.
The junction models were finite-size
with the elongation direction chosen to be along the z-axis.
The models were placed in a hexagonal supercell with $a=20~\AA$ 
and $c$ chosen to ensure that periodic replicas 
remained separated by a minimum of $10~\AA$.
The Brillouin zone sample was restricted to k=0 ($\Gamma$ only).
Constrained optimization was performed to meet a minimum force
criterion of 0.005 to 0.01 $eV/\AA$.

Several constrained approaches were employed in this study.
In the most restrictive, all degrees of freedom were constrained
and the elongation of a specific bond was considered without other relaxation.
To simulate the response of one link bond, including relaxation, 
a model consisting of one electrode and a molecule was constructed.
The back-plane of the model electrode 
and the specific link bond contact atom (\textit{e.g.}, N for an amine link group)
were constrained while all other atoms were allowed to relax.
Finally, for the full junction simulations, the back-planes of the
two electrode structures were constrained, but all other atoms were relaxed.
In each case, the potential surface was mapped by taking
finite steps in the separation defined by the constraints,
with the force criterion for relaxation being satisfied at each step.
For full junction simulations, $0.1~\AA$ was the typical step size, 
while for single interface simulations $0.05~\AA$ was used.
In selected cases, smaller steps were used to explore regions of the
potential surface where abrupt jumps in structure were encountered.
The external force required to maintain the constrained system
was calculated using a centered, finite difference approximation.
For the examples where this could be easily compared to the internally computed
forces on constrained atoms, the accuracy was found to be better than 0.05 nN,
with 0.01-0.02 nN being more typical.

\begin{figure}[b]
\centering
\includegraphics[width=3.1in]{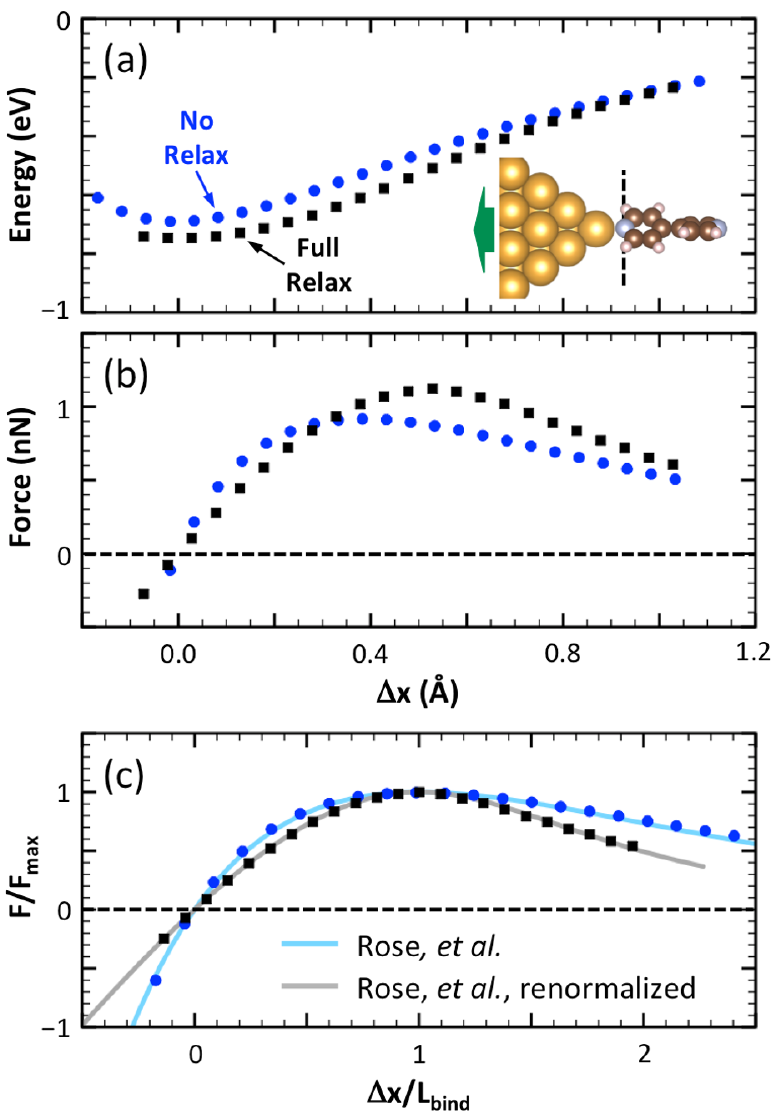}
\caption{
(a) Potential energy surface for stretching the BP bonded to a pyramidal Au tip,
as illustrated in the inset.  For full relaxation, the back-plane of the pyramid
and the bonded N atom are fixed at each step, while all other atoms relax.  
For no relaxation, all atoms are fixed at each step and only the N-Au bond distance is changed.
(b) Corresponding force extension characteristics.
(c) Scaled force extension characteristics, compared to the universal form of Rose, \textit{et al.},
and a curve that represents the renormalized Rose, \textit{et al.}, potential 
due to a series harmonic potential with $K_{eff} = 2.15$ as described in the text.
}
\label{fig:BipyBond}
\end{figure}

Application of this GGA approach to bulk Ag and Au yielded  lattice parameters 
(4.155 and 4.173 $\AA$) larger than experiment (4.09 and 4.08 $\AA$), 
typical of the accuracy reported for PBE \cite{Klimes11}.
To probe the accuracy of the metal-organic donor acceptor bond energy,
results for the $Au - NH_3$ molecular complex were compared to 
those from carefully converged coupled-cluster calculations \cite{Lambropoulos02}.
The GGA $Au-N$ bond energy was too small by 0.11-0.13 eV.
Previous analyses of the binding of ammonia and pyridine
adsorption energies on Au(111) suggested at least 0.1 eV underestimate as well \cite{Bilic02b, Bilic02a}.

The donor-acceptor organo-metallic bonds in this study
induce a local dipole near the bond.
For example, C8NH2 (1,8-diaminooctane) bonded
to the $Au_{20}$ pyramid tip induces a dipole of nearly 10 D.
The dipole for bonds to the trimer tip is less, but still significant.
As the bond is elongated, the induced dipole gets smaller.
The consequences for junction force-extension characteristics 
will be discussed in the next section.
Selected tests with larger supercells (up to twice as large)
showed that artifacts due to interactions with periodic replicas
were small for the chosen supercell size, namely less than 0.02 eV for junction formation energy
and 0.02 nN for $F_{max}$.

\section{Results and Discussion}

Consider first the Au-N bond at the
Au pyramid BP interface.
As illustrated in Fig. \ref{fig:BipyBond},
The proximal N position (dashed line) and the pyramid back-plane are held fixed.
The vertical BP remains vertical as these are systematically separated,
to probe the potential landscape of the bond.
In this geometry, the lone pair on the N should have maximal overlap
with the $z^2~d-$state on the apical Au atom, which is vertical in this case.
This should optimize the strength of the donor-acceptor bond,
calculated to be 0.75 eV.
Allowing for relaxation, $F_{max}$ = 1.12 nN occurs at $L_{bind}$ = 0.53 $\AA$.

\begin{figure}[b]
\centering
\includegraphics[width=3.1in]{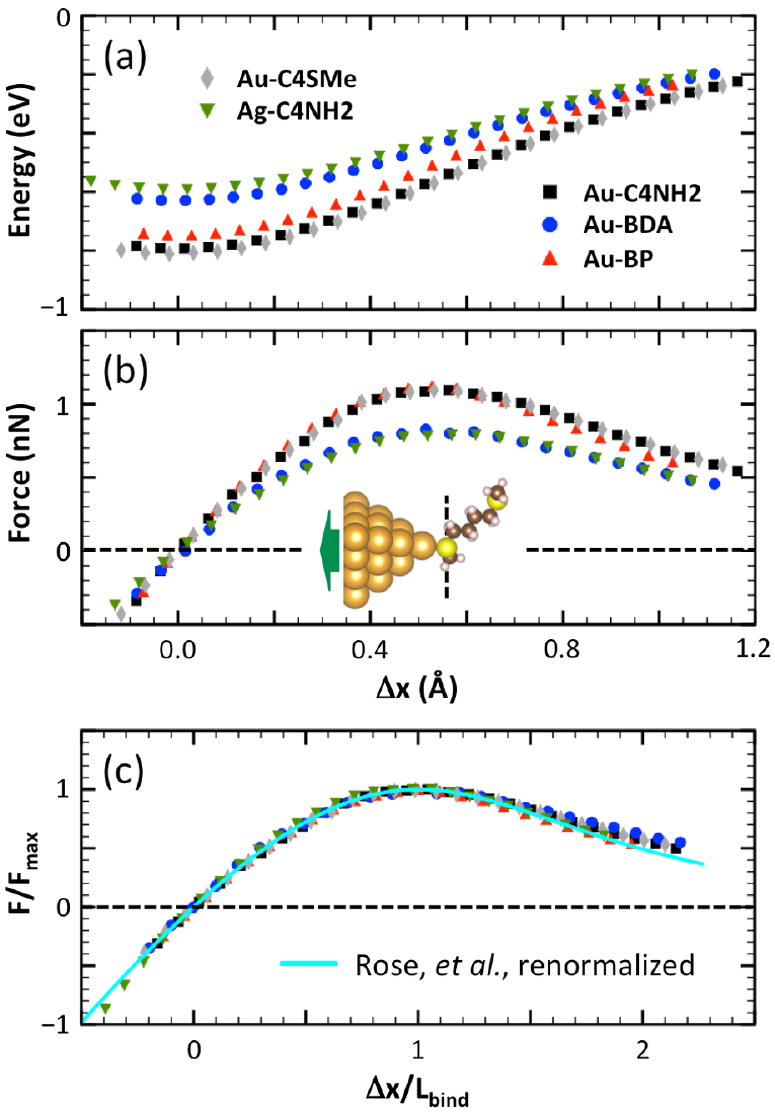}
\caption{
(a) Potential energy surfaces for stretching the interface, link-electrode bond: 
Four molecules bonded to a pyramidal Au tip and one example, C4NH2,
bonded to a pyramidal Ag tip.  See Fig. \ref{fig:ModelStructures} for constituent structures.
In all cases, the back-plane of the pyramid
and the bonded link atom are fixed at each step, while all other atoms relax.  
(b) Corresponding force extension characteristics.
Inset: visualization of the Au-C4SMe interface structure.
(c) Scaled force extension characteristics, 
compared to the renormalized Rose, \textit{et al.}, 
potential due to a series harmonic potential with $K_{eff} = 2.15$
as in Fig. \ref{fig:BipyBond}.
}
\label{fig:CompareBonds}
\end{figure}

To remove the impact of other relaxation,
while maintaining a common reference at large separation,
the BP molecule was moved in steps from beyond 1 $\AA$ separation
in towards the Au  pyramid.
Both components were frozen in their structures computed in isolation.
The resulting bond energy is calculated to be 0.69 eV,
smaller than before because the proximal structure near the energy minimum is not optimized.
Relative to the minimum energy position, the initial stiffness is larger.
At $L_{bind}$ = 0.39 $\AA$, $F_{max}$ = 0.92 nN.
As illustrated in the scaled force extension plot in Fig. \ref{fig:BipyBond}c,
the potential is well fit by the universal form of Rose, \textit{et. al}.
Furthermore, with reference to Table \ref{tab:BondModels},
the corresponding predicted $U_0 = -2.7183 \times F_{max} L_{bind}$
gives a bond energy of 0.61 eV; the same calculation for the Morse potential yields 0.65 eV.
Both are in reasonable agreement with the actual value from DFT.
 
 Now return to the full calculation with relaxation.
 The scaled force-extension plot makes it clear that this can not be fit
 by any of the simple, analytical models as such.
 However, it is natural to try the combination of a bond model
 and a series harmonic term 
\textcolor{black}{
(Eqs. (\ref{eq:HarmonicX} - \ref{eq:HarmonicL})). 
}
 Specifically, choosing $K_{eff} = 2.15$ in units of $F_{1,max}/L_{1,bind}$
 that reference the bond and recomputing a normalized force extension 
 curve from the universal model of Rose, \textit{et. al}, gives a
 good fit to the DFT-calculated data (Fig. \ref{fig:BipyBond}c).
 From Eq. (\ref{eq:HarmonicL}), the value $K_{eff} = 2.15$ implies
 $L_{bind} = 1.47 \times L_{1,bind}$.  Using the calculated values
 noted above then gives $L_{1,bind} = 0.36~\AA$ and $F_{1,max}$ = 1.12 nN.
 This value of $L_{1,bind}$ agrees well with the N-Au bond expansion
 found in the DFT calculations between equilibrium and maximum sustainable force,
 namely $0.37~\AA$.
 From this, a bond energy is inferred from $U_0 = -2.7183 \times F_{1,max} L_{1,bind}$
 to be 0.68 eV; the same calculation for the Morse potential yields 0.73 eV.
 Both agree reasonably well with the original DFT calculated value of 0.75 eV.
The fit to the renormalized model, one that accounts for
an effective harmonic potential component in series with the bond,
covers a finite section of the force extension curve including the
force maximum and some distance beyond.
These results suggest that this approach may be sufficient 
to extract the bond energy to a good approximation.

\begin{center}
\begin{table*}[t]
\centering
\begin{tabular*}{0.95\linewidth}{@{\extracolsep{\fill}} l c c c c c c}
\hline
\hline
  & \multicolumn{3}{c}{DFT} & \multicolumn{3}{c}{Renormalized Rose, \textit{et al.}} \\
Interface Bond  & $E_{bind}~(eV)$ & $F_{max}~(nN)$ & $L_{bind}~(\AA)$ & $K_{eff} ~(F_{1,max}/L_{1,bind})$ & $L_{1,bind}~(\AA)$ & $E_{1,bind}~(eV)$ \\
\hline
Au-C4NH2 & 0.79 & 1.10 & 0.54 & 2.90 & 0.40 & 0.75\\
Ag-C4NH2 & 0.59 & 0.79 & 0.58 & 2.24 & 0.40 & 0.54 \\
Au-BDA & 0.63 & 0.83 & 0.52 & 4.31 & 0.42 & 0.59 \\
Au-BP & 0.75 & 1.11 & 0.53 & 2.15 & 0.36 & 0.68 \\
Au-C4SMe & 0.81 & 1.10 & 0.54 & 3.02 & 0.40 & 0.75 \\
\hline
\hline
\end{tabular*}
\caption{
Summary of the interface bond characteristics computed for
five junctions and shown in Fig. \ref{fig:CompareBonds}.
The first set of data represent the DFT-based computations.
The second set of data results from fitting the
renormalized Rose, \textit{et al.}, potential due to a series harmonic potential
of fitted stiffness $K_{eff}$, exptessed in dimensionless units in the table,
as discussed in the text.
}
\label{tab:BondAnalysis}
\end{table*}
\end{center}

\begin{figure}[b]
\centering
\includegraphics[width=3.1in]{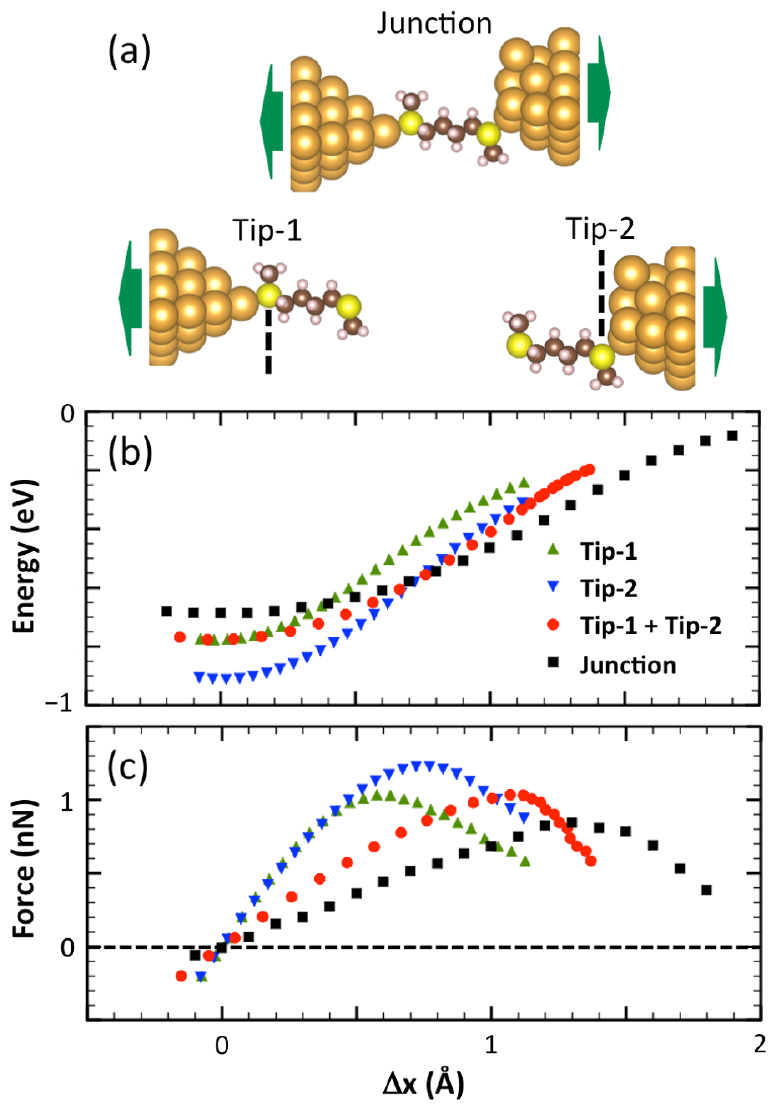}
\caption{
Model for the Au-C4SMe junction mechanics based on calculated, individual
link bond force extension characteristics.
(a) Structural model for the Au-C4SMe junction near its minimum energy configuration
together with models for the left and right link bond (Tip-1 and Tip 2) respectively.  
The link S atom is fully constrained and the opposite S atom is constrained perpendicular
to the junction axis to maintain relative orientation of the molecule during link bond elongation.
(b) Potential energy surface for each of Tip-1, Tip-2 and the Junction.
Also shown is the combination of the potential energy surfaces for Tip-1 and Tip-2
as a one dimensional mechanical model for the junction 
following Eqs. (\ref{eq:JunctionX} $-$ \ref{eq:JunctionF})
based on static equilibrium (Tip-1 + Tip-2).
(c) Corresponding force extension characteristics.
}
\label{fig:SMeComposite}
\end{figure}

To probe the generality of this picture,
DFT-calculated force extension characteristics for five different
cases are collected in Fig. \ref{fig:CompareBonds},
all with full relaxation.
These cases cover the range of molecules and link groups
covered in prior research and include the BP case previously discussed.
These model cases were all constructed so as to optimize the
donor-acceptor bond, along the lines discussed for the BP case.
However, in the cases of BDA, C4NH2 and C4SMe, the lone pair is at an obtuse
angle relative to the main backbone of the molecule.
Therefore, as illustrated previously for BDA \cite{Frei11}, the optimized orientation
of the molecule in this calculation, 
as shown in the inset to Fig. \ref{fig:CompareBonds}b for Au-C4SMe,
may differ from what would pertain
in a full junction.
Key parameters characterizing the potential energy and force extension
of these five interface bonds are gathered in Table \ref{tab:BondAnalysis}.
The data for the bond energies and maximum sustainable forces
track and form two close groupings.
The values for $L_{bind}$ all fall in the range $0.52 - 0.58~\AA$.
As the scaled force extension plot (Fig. \ref{fig:CompareBonds}c) shows,
the bond model renormalized by a single effective harmonic potential
accounts reasonably well for the data, based on the value that fit the Au-BP case.

While the deviations that are most apparant beyond the force maximum appear minor,
they raise the natural question of whether the renormalized model can be fit
to each individual trace and the extent to to which the correponding effective stiffness will vary.
The fit parameter is the value of $K_{eff}$.
Since static equilibium implies that the magnitude of the force
associated to each component of the mechanical model in series is the same,
it is easiest to use the force, derived from the DFT calculations for each value of  $\Delta x$,
as the fixed part of the data set and the $\Delta x$ as the data to be fit by the model,
through Eq. \ref{eq:HarmonicX}.
The value of $\Delta x_1$ that here refers specifically to the local bond
is obtained by inverting the analytical Rose, \textit{et al.}, model for that force, Eq. \ref{eq:RoseF}.
Then the value of $K_{eff}$ is determined by least squares fitting to the error in 
the modeled versus actual values of $\Delta x /L_{bind}$, all carried out in terms of scaled variables.
The net result of this procedure was already illustrated for Au-BP in Fig. \ref{fig:BipyBond}.

The results of this fitting procedure are shown in Table \ref{tab:BondAnalysis}.
Clearly, the variation in the value of $K_{eff}$ is of some significance.
Correspondingly, the ratio $L_{1,bind}/L_{bind}$ varies from 68\% to 81\%.
Physically, some of the inferred variations in tip stiffness can be rationalized.
For the N-Au bond in Au-BDA, this interface bond is relatively weak and 
correspondingly, there is less extension within the tip, 
thus probing a stiffer part of the Au-Au potential on average.
On the other hand, Au-BP and Au-C4NH2 have the same maximum sustainable force,
but induce a different stiffness response in the Au tip.
This may indicate the impact of different degrees of charge transfer to the Au.
But overall, the main conclusion is that fitting to a renormalized single bond model
does a generally very good job of predicting the interface bond energy.

Next I consider the junction response as a whole
and how it might differ from that predicted from
the force extension characteristics of the two link bonds computed separately.
To probe this, the Au-C4SMe junction was deconstructed.
The structure and the potential surface calculated for the full junction
are shown in Fig. \ref{fig:SMeComposite}.
From the junction structure near its equilibrium,
two model interface bonds are formed, as illustrated,
holding the orientation of the molecular backbone fixed.
Then the potential surface for each tip separately is computed
in the same way as previously described. 
The proximal S atom is fixed (dashed line) while the outer S atom is only constrained
perpendicular to the axis of elongation to maintain backbone orientation.
Then the distance between the fixed S atom and the tip base layer are separated
in steps.
Both potential surfaces are different from the one reported in Fig. \ref{fig:CompareBonds}.
The overlap between the S lone pair orbital and the contact Au d-orbitals
depends on both the angle and the local Au atomic structure near by.
As shown in Fig. \ref{fig:SMeComposite}, Tip-1 (the pyramid electrode) 
has a smaller bond energy and maximum sustainable force than Tip-2 (the trimer electrode).
The S-Au bond for Tip-1 is the bond that breaks in the full junction,
so the potential surface of Tip-1 aligns with that of the full junction near equilbrium.

From the data for the potential surface and force extension for Tip-1 and Tip-2 separately,
a mechanical model with the two components in series was constructed based on static equilibrium,
without any assumption about either component being harmonic.
This model shows the same $F_{max}$ as Tip-1, as required, but a significantly larger $L_{bind}$
and correspondingly a smaller stiffness.
This reflects the storage of energy in the stretching of Tip-2, similar to the simple model in Fig. \ref{fig:ModelJunctions}.
However, this physical effect by itself is not sufficient to explain the full junction results.

\begin{figure}[b]
\centering
\includegraphics[width=3.1in]{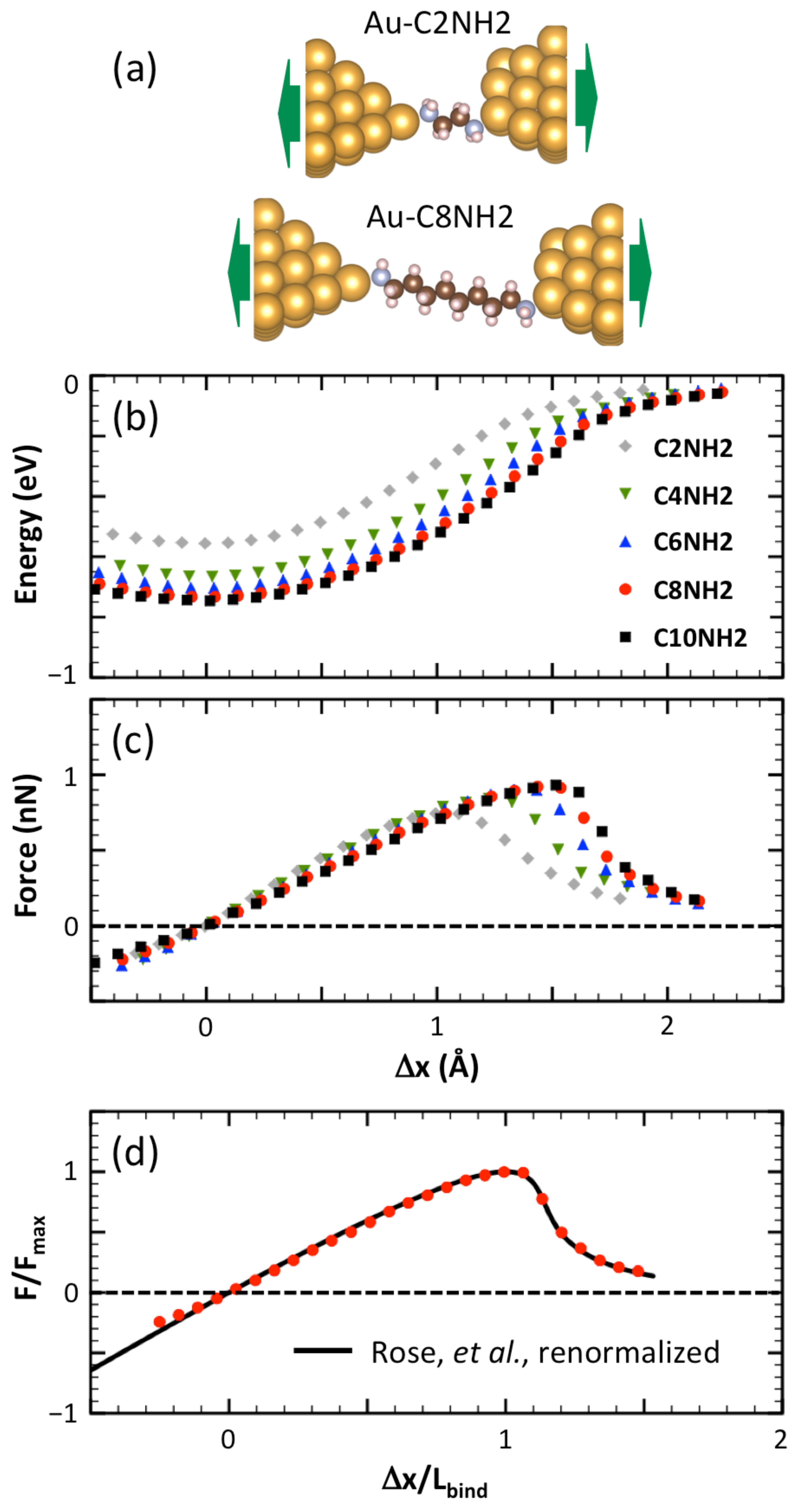}
\caption{
Comparison of the potential energy and force extension characteristics
for the amine-linked alkane junction series Au-C2NH2 to Au-C10NH2.
(a) Structural model the Au-C2NH2 and Au-C8NH2 junctions 
near their minimum energy configuration.
(b) Potential energy surfaces for the series of junctions from Au-C2NH2 to Au-C10NH2.
All junctions rupture at the pyramid tip - N bond.
In each case, the zero of energy corresponds 
to the relaxed, isolated fragments following bond rupture.
(c) Corresponding force extension characteristics.
(d) Scaled force extension characteristic for Au-C8NH2, 
compared to the renormalized Rose, \textit{et al.}, 
potential due to a series harmonic potential with
\textcolor{black}{
$K_{eff} = 0.47(F_{1,max}/L_{1,bind})$
where $F_{1,max}$ and $L_{1,bind}$ are characteristic of the N-Au bond that ruptures. 
}
}
\label{fig:AmineAlkanes}
\end{figure}

First, as compared to the model based on Tip-1 and Tip-2,
the full junction exhibits a smaller $F_{max}$ (0.85 nN versus 1.04 nN)
and a larger $L_{bind}$ ($1.27~\AA$ versus $1.07~\AA$).
Interestingly, the energy scale from the product $L_{bind} \times F_{max}$
is nearly the same (within 3 \%).
As the full junction is elongated, all the internal degrees of freedom
can fully relax.
For example, the molecule backbone does rotate a few degrees 
which alters the pathway through the individual interface bond potential surfaces.
If this type of effect were the sole difference, it is conservative
and indeed the energy parameter would be the same.

However, the second difference is in the equilbrium junction energy itself,
which is not as deep as implied by the Tip-1 bond dissociation energy by itself.
Since the Tip-1 and Tip-2 structures were specifically chosen from
the junction structure near equilibrium,
this implies that there are other interactions in play in the full junction.
I have already discussed the dipole that is induced when the interface
donor-acceptor bond is formed, \textit{e.g.}, N-Au in Sect. III.
This will result in a long-range dipole-dipole interaction across the junction
that is beyond what is captured in the Tip-1 and Tip-2 characteristics separately.
In addition, there is tunnel coupling across the junction,
which in general should lower the energy.
Both effects should be junction separation dependent.

To probe junction separation dependence,
I consider the series of diaminoalkane junctions C2NH2 to C10NH2.
This keeps the chemical characteristics of the interface bond
as fixed as possible.
The initial, relative orientation of the alkane backbone and the
alignment of the N lone pair orbital to the proximal Au atom
to which the donor-acceptor bond forms are both kept approximately fixed
in the construction of the series of junctions, 
also in an effort to make comparisons across the series as reliable as possible.
Two of the junction structures are illustrated in Fig. \ref{fig:AmineAlkanes}a.
The potential surface and force extension characteristic are shown
in Figs. \ref{fig:AmineAlkanes}b and c respectively.
The energy zero is taken from separated fragments following
rupture of the weaker of the two interface bonds (namely the bond to the pyramid tip).

Several trends emerge.
First, the shorter alkanes clearly result in junctions with a smaller net bond energy.
Second, the differences are systematically getting smaller for longer alkanes.
Third, these differences are also reflected in the stiffness of the junction
near equilibrium, with longer alkanes having a softer response.
Correspondingly, the $L_{bind}$ and the $F_{max}$ are larger for the longer alkanes.
For C2NH2, $L_{bind} = 1.03~\AA$ and $F_{max} = 0.74~nN$,
while $L_{bind}$ is $1.46~\AA$ and $1.48~\AA$,  and $F_{max}$ is 0.92 nN and 0.93 nN,
for C8NH2 and C10NH2 respectively.
Fourth, the shape of the force extension characteristic is clearly affected,
with the shorter alkanes exhibiting a more symmetrical response around
the force maximum.
However, an overall shape with a sharper drop-off after
the maximum appears to be the limit for the longer alkanes.

\begin{figure}[b]
\centering
\includegraphics[width=2.5in]{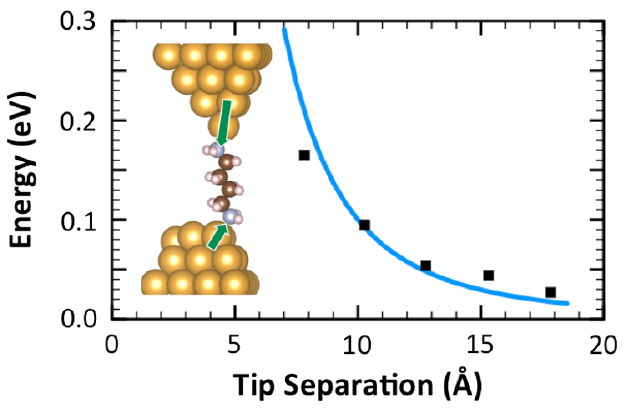}
\caption{
Energy difference between the full junction with two N-Au link bonds and
the sum of the energies of the individual bonds calculated in the absence of the other tip
for the series of Au-C2NH2 to Au-C10NH2 junctions.
This is plotted against the distance between the Au-atoms on each tip participating in the N-Au bonds.
The cooperative effects in the full junction give a net, length-dependent repulsive contribution.
The line is a guide to an estimate for dipole-dipole interactions induced by the link bonds.
Inset: Model structure for the Au-C4NH2 junction near its minimum energy configuration
with green arrows superposed to qualitatively indicate the interface dipoles
associated with the N-Au donor-acceptor link bonds.
}
\label{fig:AmineAlkaneExcessEnergy}
\end{figure}

To probe this more quantitatively,
I consider a metric for the cooperative junction effect on the binding energy.
First, define $E_{jcn,bind}$ as the energy cost to remove
the molecule completely from the junction, relative to the three resultant fragments.
This is the net binding energy of the molecule in the junction, including cooperative effects.
Second, define $E_{1,bind}$ and $E_{2,bind}$ as the energy cost to
break the bond to Au tip-1 in the absence of Au tip-2 
relative to the two resultant fragments, and the opposite.
These are the binding energies associated with each interface separately,
with no cooperative effects.
Then the net cooperative effect is:
\begin{equation}
\Delta E = E_{1,bind} + E_{2,bind} - E_{jcn,bind}.
\label{eq:CooperativeE}
\end{equation}
Correspondingly, the junction separation is measured by the
distance between the individual Au atom on each tip to which the respective
amine link group binds.
This calculation was performed starting from the junction structure near its equilibrium separation.
It can be done with all fragments frozen, or they can be relaxed in a way that constrains
remaining bonds to be representative of the junction structure.
The final results do not depend significantly on this choice
The results, including relaxation of the fragments, 
for the alkane series from C2NH2 to C10NH2 are shown
in Fig. \ref{fig:AmineAlkaneExcessEnergy}.
As already implied by Fig. \ref{fig:AmineAlkanes}b,
the net cooperative effect on binding in the junction is to destabilize the narrower junctions.

The impact of tunnel coupling is hard to estimate for these models specifically.
Qualitatively, it should be attractive (not repulsive) and it should
depend exponentially on the molecule backbone length.
On the scale of interest here, it should be neglibable for the longer alkanes studied here.

As noted in Sect. III for the Au-C8NH2 junction model specifically,
the N-Au bond to the pyramid tip induces a dipole of $2.1~e \cdot \AA$
or about 10 D with sign that corresponds to electron transfer to the Au tip.  
The bond to the trimer induces a dipole of $1.2~e \cdot \AA$,
also corresponding to electron transfer to the Au tip.
These dipoles are predominantly aligned to the axis of the junction,
as the green arrows in the inset to Fig. \ref{fig:AmineAlkaneExcessEnergy} indicate,
and oppositely directed. 
Estimates of the dipole for Au-C4NH2 are similar.
Using the approximation that the dipoles are specifically on a common axis and oppositely directed,
the interaction energy is $W_{12} = 2p_1 p_2 / d_{12}^3$ and repulsive.
With the values mentioned for the dipoles, this energy correction is plotted versus junction
separation $d_{12}$ in Fig. \ref{fig:AmineAlkaneExcessEnergy}.
Both the magnitude and the separation dependence agree with the DFT-based data
reasonably well,
particularly considering that the definition of the dipole separation is somewhat arbitrary.

The induced dipole-dipole interactions also affect the shape of the force extension characteristic.
Stretching the donor-acceptor bond reduces the dipole.
Therefore, there are two contributions to the force:
\begin{equation}
\frac {dW_{12}} {dz} = \frac {2} {d_{12}^3} \frac {d(p_1p_2)} {dz} - \frac {6p_1p_2} {d_{12}^4} \frac {d(d_{12})} {dz} .
\label{eq:DipoleF}
\end{equation}
Since the induced dipole is reduced as junction separation is increased, the first term is also negative.
Calculations show that $p_1p_2$ drop by about 25\% from the equilibrium separation to $L_{bind}$
for Au-C8NH2, so the first term is quantatively important.
For C2NH2, the estimated affect on the junction force is about $-0.1~nN$.
Furthermore, because it drops off with junction extension, it will tend to lower
the force extension characteristic more near zero and less beyond the force maximum.
This will reduce $L_{bind}$ and $F_{max}$, as well as make the characteristic
more symmetric around the maximum.
Both effects are observed in the force extension characteristics
for the short alkane cases (C2NH2 and C4NH2)
relative to those for the longer ones (C8NH2 and C10NH2).
All together, the energy estimates and the force estimates indicate that 
induced dipole-dipole interactions account for much of the variation with length
seen among the alkane series in Fig. \ref{fig:AmineAlkanes}.
Furthermore, most of the effect will be negligable for C8, C10 and longer molecules.

This suggests testing the renormalized Rose, \textit{et al.}, potential for one
of the longer molecular examples. 
Figure \ref{fig:AmineAlkanes}d shows the fit that results with 
\textcolor{black}{
$K_{eff} = 0.47(F_{1,max}/L_{1,bind})$. 
}
This is much softer than the fits for the single interface data.
However, there are now two bonds in series, the N-Au bonds, and both deformable electrodes.
From $F_{max}$ = 0.92 nN and $L_{bind}$ = 1.45 \AA, the fit implies $L_{1,bind}$ = 0.47 \AA.
This is somewhat larger than the value inferred for an optimized N-Au bond 
from the interface bond calculation with C4NH2 (see Table \ref{tab:BondAnalysis}).
However, as I noted in the analysis of Fig. \ref{fig:SMeComposite},
there is more flexibility for relaxation in the full junction and
the pathway followed locally as the N-Au bond is elongated may be different due to the junction constraints.
Thus, a softer bond here (lower $F_{max}$ and long $L_{bind}$ may be expected.
Interestingly, bond energy inferred is 0.73 eV, essentially identical to the DFT
result for breaking the junction into two fragments.

Previously, I presented a comparison of the scaled force extension characteristic
for the series of molecular junctions Au-C4NH2, Au-BDA, Au-BP and Au-C4SMe,
as well as models for the Au-Au and Ag-Ag point contact \cite{Hybertsen16}.
The data scaled remarkably well up to the force maximum.
However, after the maximum, there were significant deviations, 
attributed in part to the stored energy in the more stable of the two bonds.
The results presented here suggest that for several of those cases, there are also cooperative effects
such as induced dipole-dipole interactions that affect the shape of the force extension characteristic.
For cases where those cooperative effects are negligable, the present analysis of C8NH2 with
a renormalized bond potential is encouraging of the idea that a tractable fitting form may be achieved.
However, it will require an additional parameter, such as $K_{eff}$ to incorporate
the physical effects of stored energy in the other, deformable parts of the junction,
a step beyond the two parameter function used previously \cite{Aradhya14, Hybertsen16}.

There is another factor that needs further analysis,
as illustrated by the Au-BP case. 
It showed the sharpest drop-off after the force maximum.
Although the molecule is long enough that cooperative effects may be minimal,
it turns out that the two Au-N bonds have very similar force maxima in that Au-BP case.
That junction is an example of one I have explored numerically in which the
instability due to nearly equal extension of the two bonds up to the point of rupture was essentially realized.
While I was able to execute fits to the renormalized Rose, \textit{et al.}, potential model,
those fits were not robust.
This example points to two important factors. 
First adequate data after the force maximum likely controls the fit.
Second, there is a class of nearly symmetric junction structures in which the details
of the elongation near the maximum depend not just on the generic storage of energy,
but on the more specific, non-linear form of those deformable components.

\section{Concluding Remarks}

I have used a series of DFT-based simulations of exemplary single molecule junctions
to dissect the contributions to the overall mechanical response of such junctions to external
stress up through and beyond the extension corresponding to the maximum sustainable force.
I have focused on link bond motifs for which there is extensive experimental data.
In this study, dispersion interactions have not been explicitly considered,
but likely will play an important role in further research.
The presentation of the force extension characteristic in scaled form, $F_{max}$ force scale
and $L_{bind}$, the distance from the equilibrium bond length 
to the extension corresponding to maximum sustainable force for length,
reveals qualitative changes in those characteristics.

When the response of the interface to strain is restricted to just the extension
of the link bond itself, say N-Au, then the force-extension characteristic closely
fits the universal form proposed by Rose, \textit{et al.}
Interestingly, when the Morse potential or the Lennard-Jones potential are
scaled in the same way, their shape is almost indistinguishable over the
range of extension from near equilibrium well through the point of maximum sustainable force.
Only at larger extension does the Lennard-Jones potential appreciably depart in form.
This corresponds to the power-law decay chosen to follow the asymptotic form for dispersion interactions.
While not considered here explicitly, dispersion interactions can have a number of important impacts,
both in quantitative analysis of the bond strength 
and in the contribution of proximal, non-specific interface interactions.

Still focusing on one interface, when the other atoms proximal to the link bond
are allowed to relax, the shape of the force extension characteristic changes.
The stiffness is reduced 
and the shape in the region following the point of maximum sustainable force changes.
Physically, this can be understood in terms of additional deformable structures
in series with the link bond.
During the initial extension, up to the point of maximum sustainable force (dictated by the link bond itself here),
some of the work is stored in those other distortions.
This stretches the characteristic in this region.
After the point of the force maximum is passed, this energy is released,
slowing the net extension of the overall system.
The region beyond the force maximum is compressed.
For a practical model, this is at least approximately captured by putting a model
such as the universal model of Rose, \textit{et al.}, in series with harmonic elements.
Analysis shows that the harmonic elements can be represented by an effective spring constant
and the best fit to the data presented here requires a stiffness for the series harmonic term
that is smaller than the bond stiffness itself near the equilibrium bond length.

The force extension characteristics for the full junctions studied here show
further effects of multiple deformable structures in series.
The weakest bond controls the maximum sustainable force of the junction as a whole,
assuming basic, static equilibrium at each point of extension.
The simplest model, e.g., combining two bonds represented by the universal model of Rose, \textit{et al.},
shows the basic physical effects of stored energy in the stiffer bond,
but the resulting shape of such a model force extension does not fit well to that of the full calculations.
The additional deformation of the electrode tips alters the shape of the force extension characteristic.

To probe cooperative effects in the junctions,
I have taken exemplary cases and isolated the DFT-based predictions
for the force extension of each of the interface bonds separately, including local relaxation effects.
When these are combined to model the full junction, again relying on static equilibrium,
the results qualitatively show the reduced stiffness and increased $L_{bind}$
observed for the junction as a whole.
However, the full junction shows a larger $L_{bind}$ and a smaller $F_{max}$.
This can be understood qualitatively.
The interior junction degrees of freedom can fully respond to the externally applied stress.
The link bonds can extend following a trajectory that is locally more optimal than that captured
by the constrained trajectories of the individual interface bonds when extended separately.
Furthermore, in terms of estimating the bond energy, the product $F_{max} \times L_{bind}$
is very similar between the two calculations.

The other key difference is a smaller net binding energy in the junction,
indicating cooperative effects in the total energy of the junction.
A systematic study of the amine-linked alkane junction series C2NH2 to C10NH2
clearly revealed a strong length dependence of this energy difference.
To a first approximation, interactions between the dipoles induced by each link bond
accounted for this energy.  
Furthermore, these effects were negligible for the longer molecules, 
\textit{e.g.}, octane and decane.

One of the objectives of this analysis has been to identify key factors that will
support accurate modeling of individual, measured force-extension characteristics.
While more research is clearly needed, the present work does point to some useful
directions.
First, cooperative effects influence the mechanical characteristics of short molecules.
These may be smaller than found here when the screening of the rest of the experimental environment is included
(extended metal electrodes and solvent).
However, even in the extreme conditions here of minimal screening, the dipole derived effects
are minimal for modest length molecules (tip separations beyond about $15~\AA$).
Second, the impact of stored energy in the second junction bond and other deformable
parts of the junction which enter in series, need to be considered in further modeling.
In experiments, part of this problem can be minimized by study of asymmetrical molecular systems
in which one side will bond more strongly to the electrode, minimizing the contribution of that bond.
The other promising avenue will be further analysis of simple forms modified by an
effective harmonic representation of the other series contributions.  This does involve a third
parameter in fitting and analysis. 
Experience with fitting the simulation results suggests that the information 
contained in the region of the force extension characteristic near the force maximum and to
some distance beyond will be important.
While the target single molecule junction structures should generally be mechanically
stable along the pulling axis, in practice, measured junctions will exhibit bond rupture
due to other fluctuations and instabilities.  
This will naturally limit the scope of the force extension characteristic segment available for fitting.
It may be challenging to capture more data in the measured
force extension characteristics beyond the point of maximum sustainable force.

\section*{Acknowledgements}

I gratefully acknowledge close collaboration with L. Venkataraman throughout the
work described here as well as with several members of her research
group who have been coauthors on prior, joint publications.
This work was done using 
resources from the Center for Functional Nanomaterials, which is a 
U.S. DOE Office of Science User Facility at Brookhaven National 
Laboratory under Contract No. DE-SC0012704.

\bibliography{JunctionForceTheory}

\end{document}